\documentclass[preprint]{ccn}  

\title{Think-Aloud Reshapes Automated Cognitive Model Discovery \\ Beyond Behavior}

\author{%
  Hanbo Xie\affmark{1} \And
  Akshay K. Jagadish \affmark{3} \And
  Lan Pan \affmark{1} \AND
  Robert C. Wilson\affmark{1,2}
}

\affiliation{1}{School of Psychology, Georgia Institute of Technology}
\affiliation{2}{Center of Excellence for Computational Cognition, Georgia Institute of Techonology}
\affiliation{3}{Princeton AI Lab, Princeton University}

\emails{hanboxie1997@gatech.edu, akshay.jagadish@princeton.edu, louannapan@gmail.com, rwilson337@gatech.edu}

\begin{document}

\maketitle

\begin{abstract}
Computational cognitive models discovered using large language models have so far relied solely on behavioral data. However, it is well-known that models produced from the behavioral trajectory alone are typically under-determined. In this work, we explore the use of Think Aloud traces as an additional form of data constraint during automated model discovery. When applied to the domain of risky decision-making, we find that the models discovered with think-aloud achieve significantly improved predictive performance on held-out data. Additionally, we find that the discovered models belong to different structural classes than those discovered from behavior alone for the majority of participants (69.4\%), specifically, it shifts from Explicit comparator towards Integrated utility. These results suggest that process-level language data not only improve model fit, but also systematically reshape the structure of the discovered cognitive models, enabling the identification of mechanisms that are not recoverable from behavior alone.
\end{abstract}

\section{Introduction}

Understanding human thought processes is a central goal of cognitive science. A common approach is to construct computational models that describe the mechanisms underlying human decisions and to select and validate these models based on behavioral data. However, behavioral observations alone often leave substantial ambiguity, as different computational mechanisms can produce similar patterns of choice \citep{wilson2019ten}.

One modality that has been shown to complement behavioral trajectories in cognitive science is think-aloud reasoning traces. The think-aloud protocol records process-level data in participants by capturing their intermediate reasoning in natural language, which potentially encodes structural information about decision processes \citep{ericsson1980verbal}. While early work raised concerns about their validity and scalability, recent advances in natural language processing and large language models (LLMs) have enabled systematic analysis of think-aloud traces, leading to renewed interest in linking verbal reports to the underlying cognitive process \citep{xie2025rethinking,wurgaft2025scaling,zhang2025linking,xie2024strategic}.

However, prior work in this space has mostly focused on either validating the reliability of think-aloud data or developing methods to analyze it. In contrast, a more fundamental question remains largely unexplored: \emph{Do think-aloud reasoning traces facilitate the discovery of computational structures that cannot be identified from behavior alone?} 

To address this question, we compare the models discovered from behavioral data alone with those derived using behavioral and thought-aloud data jointly, within the automatic model discovery framework \citep{rmus2025generating}. We find that incorporating think-aloud leads to significant improvements in the quality of the models produced, in terms of predictive performance, while producing qualitatively different computational models, suggesting that process-level language data systemically reshapes the structure of the discovered models.

\section{Results}

\subsection{Think-aloud improves automated model discovery outcomes}

In this work, we use an automatic model discovery framework \citep{rmus2025generating}, called \textsc{GeCCo}, in which an LLM (LLaMA-3.1-70B) iteratively generates candidate computational models as executable functions mapping task inputs to choices and evaluates them based on their fit to held-out data. At each step, the current best-performing model is provided as a reference, and the LLM is prompted to propose alternative model structures that improve upon it.

We apply this framework to a risky decision-making dataset with think-aloud reports ($N=72$), where participants verbalize their reasoning before making binary choices across 19 trials based on \citet{kahneman19790prospect}. 

\begin{figure*}[ht]
    \centering
    \includegraphics[width=0.93\textwidth]{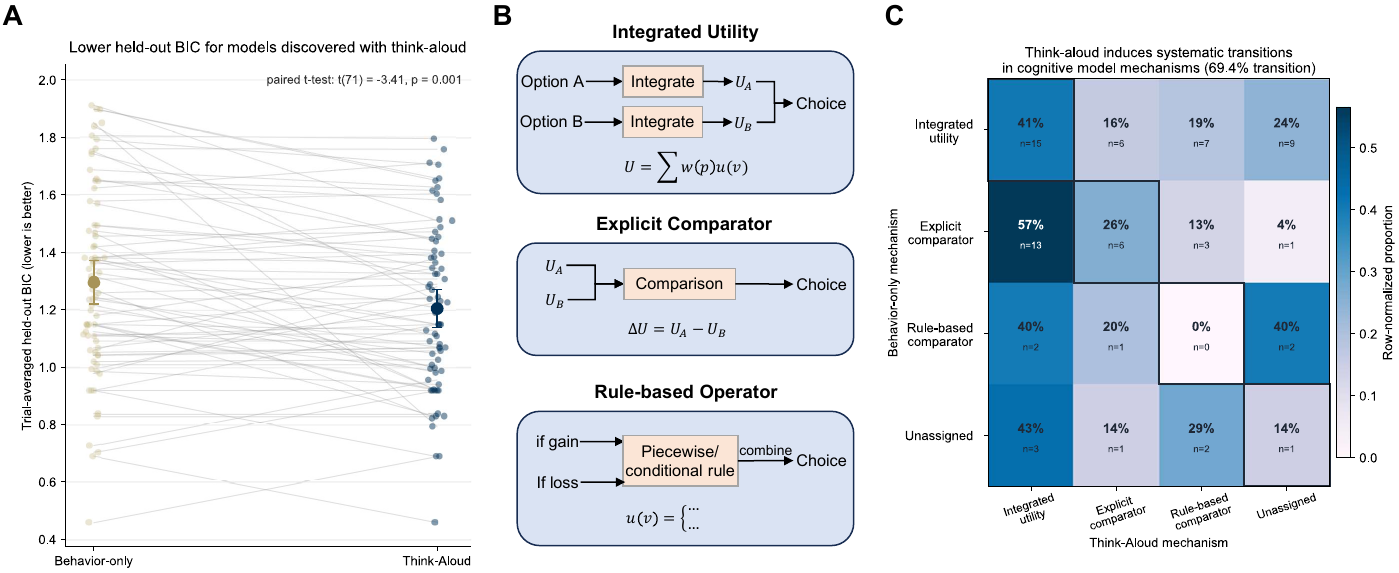}
    \caption{
    \textbf{Think-aloud improves model discovery and induces systematic shifts in discovered mechanisms.}
    \textbf{A}, Trial-averaged held-out BIC for each participant’s best discovered model under the behavior-only and think-aloud conditions. Lower BIC indicates better out-of-sample fit. Each pair of points is connected within participant; larger points show the group mean $\pm$ 95\% CI.
    \textbf{B}, Schematic definitions of the three main mechanism families identified from normalized computation graphs: \textit{Integrated utility}, which transforms and integrates each option before comparison; \textit{Explicit comparator}, which computes utilities and compares them directly (e.g., $\Delta U = U_A - U_B$); and \textit{Rule-based operator}, which applies piecewise or conditional rules before combining information into a choice.
    \textbf{C}, Row-normalized transition matrix from the behavior-only best-model cluster to the think-aloud best-model cluster. Numbers indicate proportions (counts shown below). Off-diagonal mass indicates mechanism shifts, with 69.4\% of participants transitioning to a different cluster.
    }
    \label{fig:main_figure}
\end{figure*}

We compare two conditions: models discovered using behavioral data alone, and models discovered using both behavioral and think-aloud data as input to the LLM. In both cases, candidate models are evaluated on 10 held-out trials that were not included in the prompt used during model generation. For each participant, we repeat the discovery process 10 times with the same data splits and report the best-fitting model based on Bayesian Information Criterion (BIC) \cite{watanabe2013widely}.

Among the 72 participants, 59.7\% showed lower held-out BIC under the think-aloud condition. A paired t-test confirmed this difference, showing that models discovered with think-aloud data have significantly lower held-out BIC than those based on behavior alone ($t(71) = -3.41$, $p = 0.001$), indicating improved out-of-sample model fit (Figure~\ref{fig:main_figure}A).

\subsection{Think-aloud traces reshapes the structure of discovered models}

To characterize model structure, we convert each discovered program into a normalized computation graph, extract structural features, and cluster them using HDBSCAN \citep{mcinnes2017hdbscan}. This yields three major mechanism families—\textit{Integrated utility}, \textit{Explicit comparator}, and \textit{Rule-based operator}—plus a small set of unassigned models (Figure~\ref{fig:main_figure}B). We then assign each participant’s best model under each condition to a cluster and compute a row-normalized transition matrix (behavior $\rightarrow$ think-aloud).

Upon clustering, we examine whether incorporating think-aloud data changes the structure of model identified by the discovery process. The resulting transition matrix (Figure~\ref{fig:main_figure}C) shows substantial off-diagonal mass: 69.4\% of participants are assigned to different mechanism clusters when models are discovered with versus without think-aloud data.

These transitions are not uniform: for example, models in the \textit{Explicit comparator} cluster frequently shift to \textit{Integrated utility} (57\%), while other clusters exhibit more distributed transition patterns. These transitions correspond to concrete changes in computational organization rather than superficial code variation. For example, some participants shift from models that explicitly compare option values to models that first transform and integrate gains, losses, or probabilities within each option before comparison; others show the reverse pattern. This indicates that think-aloud data can redirect discovery toward qualitatively different mechanism families (Figure~\ref{fig:main_figure}B,C).

Importantly, these two effects—improved model fit and shifts in mechanism clusters—are closely linked. Behavioral data alone often underdetermines the space of computational models, as multiple structures can produce similar choice patterns. The inclusion of think-aloud data reduces this indeterminacy by providing additional constraints on the cognitive mechanism that is underlying the observed behavior and how it is structured, thereby favoring different classes of explanations.

Together, these results show that incorporating think-aloud data does not merely improve model fit but can systematically alter the computational structures identified by automated model discovery frameworks. This suggests that process-level language data plays a functional role in constraining the space of admissible models, enabling the identification of alternative mechanisms that are not recoverable from behavior alone.

\printbibliography

\end{document}